\documentstyle[twocolumn,prl,aps]{revtex} %%%%%%%%%%%%%%%%%%%%%%%%%%%% 
\begin{document}
\draft
%\newcommand{\lapprox}{\stackrel{<}{\scriptstyle \sim}} 
%\newcommand{\gapprox}{\stackrel{>}{\scriptstyle \sim}} 
%************************************************* 
%************** D E F I N I T I O N S ************ 
%**************                       ************ 
\def\half{{1\over 2}} 
\def\vopo{(VO)$_2$P$_2$O$_7$\ } 
\def\vodpo{VODPO$_4 \cdot 1/2$D$_2$O\ } 
\def\vohpo{VOHPO$_4 \cdot  1/2$H$_2$O\ } 
\newcommand{\lapprox}{\stackrel{<}{\scriptstyle \sim}} 
\newcommand{\gapprox}{\stackrel{>}{\scriptstyle \sim}}

\twocolumn
[\hsize\textwidth\columnwidth\hsize\csname @twocolumnfalse\endcsname

%\date{\today}
%\documentstyle[12pt,aps,epsf]{revtex}
%\setlength{\footheight}{0.cm}
%\setlength{\textwidth}{16.0cm}%{19.8cm} %\setlength{\textheight}{23.0cm}
%{27.5cm} 
%\setlength{\fboxsep}{2mm} %\pagestyle{plain}

%\small \rm \begin{flushright} 
%\small{ORNL-CTP-96-08}  \\
%\small{cond-mat/9607059}\\
%\end{flushright} 
%\vspace{2cm}  
%\large \bf 
\title{\bf Magnetic Excitations in the S$=1/2$ 
Alternating Chain Compound \vopo}

\author{
A.W. Garrett$^1$, S.E. Nagler$^2$, D.A. Tennant$^2$, B.C. Sales$^2$ 
and T. Barnes$^{2,3}$ }

\address{
$^1$Department of Physics,
University of Florida,
Gainesville, FL 32611-0448\\
$^2$Oak Ridge National Laboratory, Oak Ridge, TN 37831-6393  \\  
$^3$Department of Physics and Astronomy, University of Tennessee, 
Knoxville, TN 37996-1501 \\
}

%\date{\today}
\maketitle

\begin{abstract}
Magnetic excitations in an array of \vopo single crystals have been 
measured using 
inelastic neutron scattering. Until now, \vopo has been thought 
of as a two-leg 
antiferromagnetic Heisenberg spin ladder with chains running in 
the $a$-direction.  
The present results show unequivocally that \vopo is best 
described as an alternating 
spin-chain directed along the crystallographic $b$-direction. 
In addition to the expected magnon with magnetic zone-center 
energy gap $\Delta = 3.1$ meV, 
a second excitation is observed at an energy just below $2\Delta$.  
The higher mode may be a triplet two-magnon bound state.
\end{abstract}


]
\vspace{0.3cm}

Spin gaps are found in many low dimensional interacting quantum spin 
systems.  Examples of recent interest 
include integer-spin Heisenberg antiferromagnetic chains\cite{haldane}, 
spin-Peierls systems\cite{nishi} and spin ladders\cite{dagotto,cava}. 
The insulating magnetic salt \vopo (VOPO) has been widely 
considered to be an excellent 
realization of the two-leg 
antiferromagnetic Heisenberg spin ladder\cite{jjgj}.
In this Letter we report results from
inelastic neutron scattering experiments using an aligned array 
of VOPO single crystals.  
Our results show conclusively that the magnetic properties of 
VOPO are not those of a 
spin ladder.
VOPO is instead found to be an 
alternating Heisenberg antiferromagnetic chain, with weak 
ferromagnetic interchain couplings. 
Moreover, the magnetic chains in VOPO run perpendicular to 
the supposed ladder direction.
The measured alternation parameter  describing VOPO is in 
the interesting regime in which 
a novel two-magnon bound state may occur \cite{uhrig}.

The crystal structure of VOPO is nearly 
orthorhombic, with a slight monoclinic distortion so that the 
space group is P2$_{1}$\cite{nhs}. 
The room temperature lattice parameters 
are $a$=7.73\AA, $b$=16.59\AA, $c$=9.58\AA, 
and $\beta$=89.98$^{\circ}$.
The magnetic properties of VOPO arise from $S=1/2$ V$^{4+}$ 
ions situated within 
distorted VO$_6$ octahedra. 
Face-sharing pairs of VO$_6$ octahedra are stacked in two-leg 
ladder structures 
oriented along the $a$-axis. 
The ladders are separated by large, 
covalently bonded PO$_4$ complexes. The V-V nearest neighbor 
distance in the $a$ 
direction is approximately $3.86$\AA, one half of a lattice spacing.
Along the $b$ (``rung'') direction the mean V-V distance 
is $3.2$\AA \ within the ladder.  
The inter-ladder V-V distance across the PO$_4$ 
groups is $5.1$\AA. The structure is illustrated
schematically in figure 1.

The susceptibility of VOPO powder was investigated by 
Johnston {\it et al.}\cite{jjgj}. 
Although the susceptibility could be accurately reproduced 
by either a spin 
ladder (with $J_\| \approx J_\perp$) or by an alternating 
chain \cite{jjgj,br}, the 
expectation that the PO$_4$ group would provide a weak 
superexchange path led to a 
preference for the spin ladder interpretation of VOPO. 
The proposed spin ladder is 
illustrated schematically in figure 1.  Motivated by 
the detailed theoretical 
predictions for the excitation spectrum of a spin ladder \cite{br}, 
Eccleston {\it et al.} \cite{ebbj} used pulsed inelastic neutron scattering 
to probe the dynamic magnetic properties of VOPO powders.
Their measurement of a spin gap of 3.7 meV 
was interpreted as further support for the ladder model.

The surprising discovery of a second spin excitation near 6 meV in a recent 
triple-axis neutron scattering experiment \cite{gnbs} on VOPO powder is 
inconsistent with a simple spin ladder model.
It has long been known 
that measurements of static magnetic properties cannot easily distinguish 
between spin ladder and alternating chain models \cite{cuno3}, and also that 
measurements on powders 
can be difficult to
interpret.  For these reasons we undertook 
studies of spin dynamics in VOPO single crystals.  

% EXPERIMENTAL DETAILS

Single crystals of \vopo were grown using a method described 
previously\cite{gnbs}.
To achieve sufficient sample volume for inelastic neutron 
scattering measurements, 
approximately 200 single crystals of typical size 1x1x0.25 mm$^3$ 
were aligned, 
mounted on thin aluminum plates, and assembled into an array.  
The resulting sample had an effective 
mosaic spread of  $8-10^\circ$ FWHM.  
Although not optimal, this
proved to be adequate for the present purpose.

Inelastic neutron scattering measurements were carried out using triple 
axis spectrometers at the HFIR reactor, 
Oak Ridge National Laboratory. 
The different instrumental configurations utilized are 
summarized in Table I.
For temperature control the VOPO array was mounted either in a closed cycle 
refrigerator or in a flow cryostat. Both $(h,k,0)$ and $(0,k,l)$
scattering planes were used in different experiments.

Typical constant-$\vec Q$ scans are illustrated in figure 2.
The left panels show measurements at $(1,-2,0)$ at 
temperatures of 10K (lower panel) 
and 30K (upper panel).
The 10K scan clearly shows two peaks, and the reduced 
intensity at the higher 
temperature is indicative of their magnetic origin. 
The right panel
shows a 10K scan at $\vec{Q}=(0,-2,0)$, where the 
peaks are  
slightly lower in energy and are much more intense than at $(1,-2,0)$.

The solid lines through the 10K data are obtained by 
a least-squares fit of the data to a convolution of 
the full 
instrumental resolution with delta function excitations using 
dispersion relations discussed below. At low temperatures the peak
widths are resolution limited. The significant intensity difference 
between (1,-2,0) and (0,-2,0) is seen to be a consequence of
resolution combined with the large mosaic spread of the sample. 
Simple gaussian fits to obtain the peak positions gave 
the same results as full convolutions. The peak positions were 
determined at various wavevectors.

Figure 3 shows the measured dispersion along the crystal axis directions 
indicated. 
Several features are immediately apparent. The excitation energies 
have at most a 
very weak dependence on the $Q_{c}$ component of the wavevector 
(middle panel).  
The dependence of energy on $Q_{a}$ 
(in the ``ladder'' direction) is much 
weaker than on $Q_{b}$ (in the ``rung'' direction). 
This implies that the magnetic coupling which supports spin waves 
is indeed dominantly 
one-dimensional, but that the strongest coupling is along the $b$ direction! 
Moreover, the dispersion shows that the 
exchange in the ``ladder'' direction 
is {\it ferromagnetic} in nature.

The observed spectrum of magnetic excitations in VOPO is clearly inconsistent with a 
two-leg spin ladder.  A better model must account for the strong
dispersion in the $b$ direction, the presence of a spin gap, 
and be compatible with
the crystal structure of VOPO.  The most likely model is that 
of weakly coupled
antiferromagnetic alternating chains running in the $b$ direction.  
The gap 
is a consequence of the spin dimerization inherent in the VOPO 
crystal structure.

A correct theory for VOPO must also explain the presence of two 
magnetic modes.   
Since the high temperature limit
of the magnetic susceptibility\cite{jjgj} is exactly that 
expected for $S=1/2$, g=2 the 
possibility that the upper mode is a low lying single ion 
excitation can be ruled out.

One potential explanation for a second mode is splitting 
of a triplet magnon level by anisotropy.  
Using the three
inequivalent exchanges $J_{1}$, $J_{2}$, and $J_{a}$
shown in figure 1 the Hamiltonian can be written: 
\begin{eqnarray}
\hat{H} = \sum_{l,m} \sum_{\alpha =x,y,z}
\{J_{a}^{\alpha\alpha} S^\alpha_{l,m} \, S^\alpha_{l+1,m}+ \nonumber \\
J_1^{\alpha\alpha} S^\alpha_{l,2m-1} \, S^\alpha_{l,2m} +
J_2^{\alpha\alpha} S^\alpha_{l,2m} \, S^\alpha_{l,2m+1} \}
\end{eqnarray}
where the index $l(m)$ runs over spin sites in the $\vec{a}(\vec{b})$ direction.
The anisotropic exchange is represented by diagonal matrices 
$J^{\alpha \alpha}_{\nu}= J_{\nu} [\delta^{\alpha z}+
\epsilon_{\nu}(\delta^{\alpha x}+\delta^{\alpha y})]$ 
for $\nu=1,2,a$.  Here $z$ is the unique axial
direction. 

The excitation energies can be calculated using a pseudo-boson 
technique appropriate for dimerized chains\cite{clt}.
To first order this agrees with perturbation 
theory from the isolated dimer limit\cite{abh}.
In this approximation the anisotropy leads to 
an $S^{z}=0$ longitudinal (L) mode and a
doubly degenerate $S^{z}=\pm 1$ transverse (T) mode:
\begin{eqnarray}
\omega^{L}_{\vec{Q}} & = & \{\epsilon_1 J_{1}[\epsilon_{1} J_{1}-J_{2} 
cos(\pi Q_{b})+ 2 J_{a}\cos(\pi Q_{a})]\}^{\half} \nonumber \\ 
\omega^{T}_{\vec{Q}} & = & \{ J_{1}(1+\epsilon_{1})/2 
[J_{1}(1+\epsilon_{1})/{2}-  \nonumber \\ 
& & \epsilon_{2}J_{2}\cos(\pi Q_{b})
+ 2 \epsilon_{a}J_{a}\cos(\pi Q_{a})] \}^{\half}.
\end{eqnarray}
The solid lines in figure (3) are the result of a least-squares fit 
of equation (2) to the data. 
Two equally good descriptions of
the data are possible depending on whether the upper mode (A) 
or lower mode (B) is 
chosen as the longitudinal branch; the results are 
summarized in Table II.

In principle, the $\vec{Q}$ 
dependence of the mode 
intensity should allow one to distinguish between the L and 
T modes if the model is applicable.  
In the present experiment this is complicated 
by the large 
effective sample mosaic which dominates the observed changes in mode 
intensity as the direction of $\vec{Q}$ is varied.
Nevertheless, it can be stated that both modes appear with similar
intensities for all directions measured.
This is unexpected in the anisotropic exchange model.

Fits to the observed dispersion require strong V-V   
exchange through the PO$_{4}$ groups.
A similarly strong exchange mediated by a PO$_{4}$ complex has recently 
been found in the 
chemical precursor 
compound of VOPO, \vodpo \cite{vodpo_paper}, which contains magnetically 
isolated $S=1/2$ V-V dimers.
Relatively strong superexchange through the PO$_{4}$ groups is 
possible because of 
coherent molecular electron orbitals in the complex.
Beltr\'an-Porter {\it et al.}\cite{euro_chemists} anticipated the 
importance of 
superexchange through PO$_{4}$ in VOPO and proposed an alternating 
chain in 
the $b$-direction, which is in agreement with the present result. 

Although equation (2) gives an excellent account of the data, 
considerable 
exchange anisotropy is necessary to account for the mode 
splitting. Recent single crystal magnetic 
susceptibility 
measurements \cite{thompson} were quantitatively consistent with the previous 
powder results\cite{jjgj} and found little, if any, evidence for anisotropy. 
Further, the 
exchange coupling in the precursor compound \vodpo was found to be 
isotropic\cite{vodpo_paper}. These facts suggest that one should 
seek another explanation for the second mode.

% Uhrig & Schulz model - bound two-magnon mode.
In a recent paper \cite{uhrig}, Uhrig and Schulz have considered 
the excitations of an alternating $S=1/2$ Heisenberg chain with 
isotropic exchange (equation (1) with $\epsilon_{1}=\epsilon_{2}=1$ 
and $J_a=0$) in the 
continuum limit using field theoretical methods. 
The basic excitations are triplet magnons with 
an energy gap $\Delta$.
The magnitude of $\Delta$ depends on the alternation 
parameter $\delta=(\beta-1)/(\beta+1)$ 
where $\beta=J_{1}/J_{2}$.  The limit $\delta=0$ corresponds to the usual 
antiferromagnetic Heisenberg chain for which the excitations are free 
spinons resulting in the well known continuum excitation 
spectrum\cite{kcuf3}. 
For $\delta > 0$ a gap opens in the spectrum\cite{uhrig,tsvelik} 
and the lowest lying magnetic 
excitation is a well defined $S=1$ triplet mode.  For small $\delta$ one 
expects a second gap to the remnant of the original continuum 
at an energy of $2\Delta$.  
Both the continuum\cite{arai} and second gap\cite{ain} have been 
reported in CuGeO$_{3}$, 
for which $\delta \lapprox 0.05$. 
The theory further predicts that for $\delta \gapprox 0.082$ a 
second sharp triplet 
mode appears at energies just below the continuum\cite{uhrig}.  
This mode is a two-magnon bound state.  Assuming isotropic exchange
the alternation parameter describing VOPO 
is $\delta \approx 0.1$, in the regime where this mode should be visible. 
Given the 
evidence in favor of isotropic exchange in VOPO, it is plausible 
that the observed 
second mode is in fact the two-magnon bound state.

Figure 4 shows an expanded
view of the dispersion in VOPO along the chain direction 
near $\vec{Q}=(0,-2,0)$, which 
corresponds to $Q=\pi$ in reduced units.
The solid and dashed lines are the result of fitting the 
one-dimensional lower ($l$) 
and upper ($u$) mode dispersions to the expression 
\begin{equation}
\omega^{l,u}_{Q}=\sqrt{(\frac{\pi}{2}J)^{2}\sin^{2}(Q)+\Delta^{2}_{l,u}} 
\end{equation}
resulting in $J=9.3(1)$ meV, $\Delta_{l}=3.12(3)$ meV, 
and $\Delta_{u}=5.75(2)$ meV.  The 
expression for the lower mode in equation (3)
agrees with the predictions of the continuum
model\cite{uhrig} for $Q \approx \pi$. 
The value of $\Delta_{l}/J$ allows a determination of $\delta$
using the results of numerical computations\cite{uhrig}.
Note that $J=(J_{1}+J_{2})/2$.
The value of $\Delta_{u}$ is slightly 
less than $2\Delta_{l}$ as expected for 
the two-magnon bound state. In contrast, 
if the second mode is a consequence of anisotropy 
this proximity is purely accidental.
We believe that a two-magnon bound state is the best 
explanation for the upper magnetic mode. 

Several open questions about this interpretation remain.
The substantial intensity of the upper mode is very unusual
for two-magnon scattering.
As indicated schematically in figure 4 there may also be a 
higher lying continuum of excitations 
but no clear sign of this has appeared in our measurements to date.  
Detailed theoretical 
calculations of the expected neutron scattering response 
would be of great interest. Light
scattering measurements may be sensitive to
two-magnon states and should also be instructive.

In the presence of a sufficiently strong external magnetic field the 
predicted mode splittings 
of the two proposed models are very different.
Unfortunately such experiments are extremely difficult 
with the VOPO samples available to 
date.  A suitably large {\it single} crystal is necessary for 
this as well as for
careful investigations of the mode intensities as a 
function of $Q$ and T. Continuing efforts are 
underway to obtain the desired sample.

We would like to acknowledge useful interactions with P. Dai, J. Fernandez-Baca,
H. Mook, and J. Zarestky.
We thank J. Thompson for sharing results of 
susceptibility measurements with us prior to publication.
Expert technical assistance was provided by Scott Moore and Brent Taylor.
This work was supported in part by the United States Department of 
Energy under contract 
DE-FG05-96ER45280 at the University of Florida, and by Oak Ridge 
National Laboratory, 
managed for the U.S. D.O.E.
by Lockheed Martin Energy Research Corporation under 
contract DE-AC05-96OR22464.
\begin{center}
\begin{table}
\caption{Spectrometer configurations$^{\dag}$}
\begin{tabular}{lcc}
Instrument & Collimation & Energy \\
\tableline
HB1A & $40'-40'-40'-68'$ & $E_I$=14.7meV \\ 
HB1 & $48'-40'-60'-120'$ & $E_F$=13.5meV \\ 
HB3 & $48'-40'-60'-120'$ & $E_I,E_F$=30.5meV \\ 
\end{tabular}
$^{\dag}$Using PG(002) monochromator, analyzer, PG filter.\\
\end{table}
\end{center}

\begin{center}
\begin{table}
\caption{Anisotropic exchange parameters} 
\begin{tabular}{lcc} Parameter & fit A  & fit B \\
\tableline
$J_a$ (meV) & $-0.27(2)$ & $-0.24(2)$ \\
$\epsilon_a$ & $0.91(7)$ & $1.13(9)$ \\ 
$J_1$ (meV) & $10.30(14)$ & $12.32(55)$ \\ 
$\epsilon_1$ & $1.16(3)$ & $0.93(5)$ \\ 
$J_2$ (meV)& $8.7(3)$ & $10.18(15)$ \\ 
$\epsilon_2$ & $1.12(2)$ & $0.85(4)$ \\ 
\end{tabular} \end{table}
\end{center}

\newpage

\begin{center}
{\Large Figure Captions}
\end{center}

\begin{figure}
{Figure~1 (color).
Schematic depiction of the structure and magnetic interactions in VOPO.  
The spin ladder model previously thought to describe VOPO has 
nearest neighbor exchange constants $J_{\parallel}$ along 
the $a$ (``ladder'') direction and $J_{\perp}$ along 
the $b$ (``rung'') direction.  In the alternating chain model, 
nearest neighbor V$^{4+}$ ions are alternately coupled by 
constants $J_{1}$ and $J_{2}$ along the $b$ (chain) direction.  
Neighboring spins in adjacent chains are coupled by $J_{a}$.  
Magnetic coupling in the $c$ direction is negligible.
}
\end{figure}

\begin{figure}
{Figure~2.
Typical constant-$\vec Q$ scans at $\vec{Q}=(1,-2,0)$ (left panels) 
and $\vec{Q}=(0,-2,0)$ (right panel).  
The solid lines running through the 10K data are least-square fits of a 
convolution of the instrumental resolution with delta function 
excitations obeying the dispersion relations of equation (2).  
The solid line through the 30K data is a guide to the eye. Intensities
are normalized to a fixed number of incident neutrons expressed by
monitor count units (MCU). Some representative $1\sigma$ error bars
are shown.
}
\end{figure}

\begin{figure}
{Figure~3.
Measured dispersion of magnetic excitations in VOPO at T = 10K.  
When not visible error bars are smaller than the size of the plotted symbols.  
Filled circles (open diamonds) are points from the lower (upper) energy mode.
The solid lines are dispersion curves calculated using 
parameters obtained by fitting equation (2) to 
the observed peak positions. Wavevectors are plotted in units 
corresponding to the VOPO reciprocal lattice.
}
\end{figure}

\begin{figure}
{Figure~4.
Closeup of the measured dispersion in VOPO in the chain 
direction near the antiferromagnetic zone center (0,-2,0), 
($Q=\pi$ in reduced units).  The lines through the upper and lower 
modes are from a fit of equation (3) to the data for 
wavevectors along the chain direction alone.
The hypothesized continuum is illustrated schematically.
}
\end{figure}

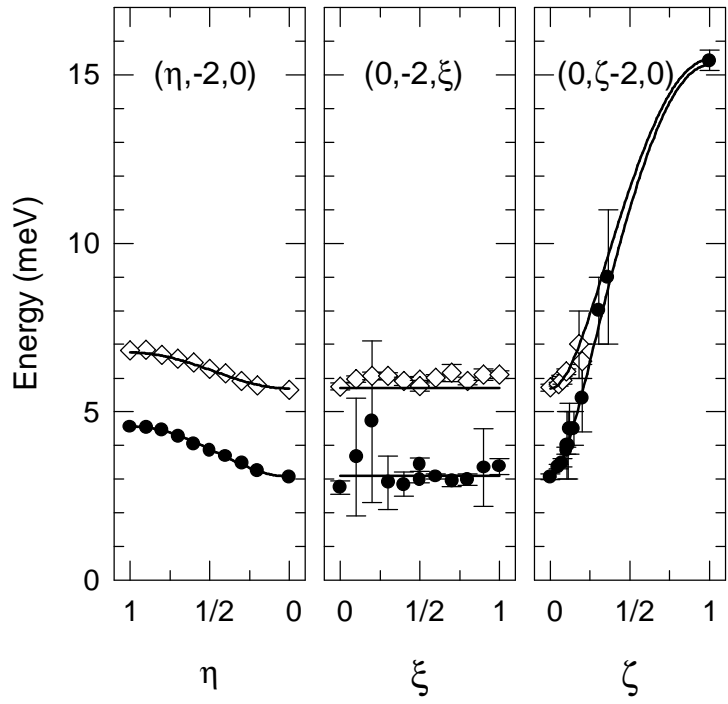
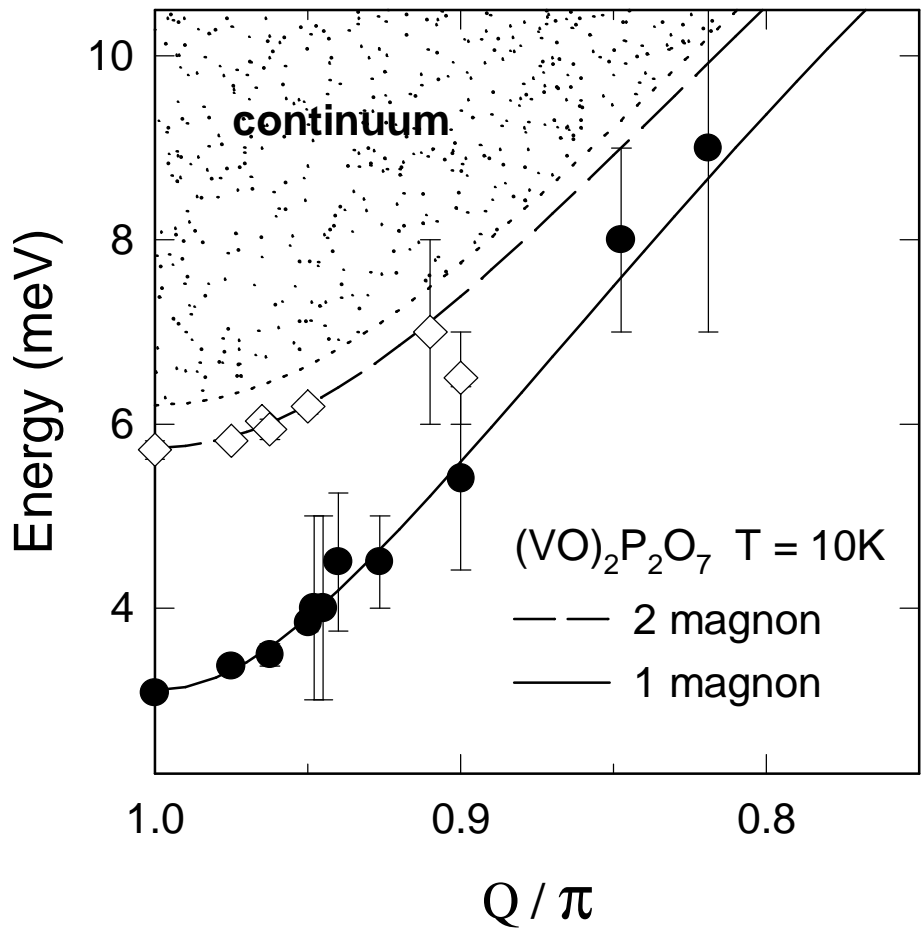
\end{document}